# RENiO$_3$ single crystals (RE = Nd, Sm, Gd, Dy, Y, Ho, Er, Lu) grown from molten salts under 2000 bar oxygen-gas pressure

Yannick Maximilian Klein,*[a] Mirosław Kozłowski,[b] Anthony Linden,[c] Philippe Lacorre,[d] Marisa Medarde*[a] and Dariusz Jakub Gawryluk*[a]

a. Laboratory for Multiscale Materials Experiments, Paul Scherrer Institut, Forschungsstrasse 111, CH-5232 Villigen PSI, Switzerland.

b. Łukasiewicz Research Network - Tele & Radio Research Institute, 11 Ratuszowa Street,03-450 Warsaw, Poland

c. Department of Chemistry, University of Zürich, Winterthurerstrasse 190, 8057 Zürich, Switzerland

d. Institut des Molécules et Matériaux du Mans (IMMM) - UMR 6283 CNRS, Le Mans Université, Avenue Olivier Messiaen, 72085 Le Mans, France

Corresponding author Emails: maximilian.klein@psi.ch; marisa.medarde@psi.ch; dariusz.gawryluk@psi.ch

**ABSTRACT:** The electronic properties of transition-metal oxides with highly correlated electrons are of central importance in modern condensed matter physics and chemistry, both for their fundamental scientific interest, and for their potential for advanced electronic applications. The design of materials with tailored properties has been, however, restricted by the limited understanding of their structure-property relationships, which are particularly complex in the proximity of the regime where localized electrons become gradually mobile. RENiO$_3$ perovskites, characterized by the presence of spontaneous metal-to-insulator transitions, are one of the most widely used model materials for the investigation of this region in theoretical studies. However, crucial experimental information needed to validate theoretical predictions is still lacking due to their challenging high-pressure synthesis, which has prevented to date the growth of sizable bulk single crystals with RE ≠ La, Pr and Nd. Here we report the first successful growth of single crystals with RE =Nd, Sm, Gd, Dy, Y, Ho, Er and Lu and sizes up to ~75 μm, grown from molten salts in temperature gradient under 2000 bar oxygen gas pressure. The crystals display regular prismatic shapes with flat facets, and their crystal structures, metal-insulator and antiferromagnetic order transition temperatures are in excellent agreement with previously reported values obtained from polycrystalline samples. The availability of such crystals opens access to measurements that have hitherto been impossible to conduct. This should contribute to a better understanding of the fascinating properties of materials with higly correlated electrons, and guide future efforts to engineer transition metal oxides with tailored functional properties.

INTRODUCTION

The properties of transition-metal oxides (TMOs) with correlated electrons at the boundary between localized and itinerant behaviour have fascinated the scientific community since the pioneering work of Mott in the early 1960's.[1] High temperature superconductivity, colossal magnetoresistance and metal-to-insulator transitions (MITs) are prominent examples of the exotic properties that can emerge in TMOs by approaching this region. Given their huge potential for applications in optoelectronics, data storage, neuromorphic and quantum computing technologies, understanding the origin of these properties is of fundamental importance for material design. However, the comprehension of the structure-property relationships in this kind of materials is less developed than in conventional semiconductors, where the electronic structure is well described by density-functional band theory and many functional characteristics such as the band-gap can now routinely be engineered.[2] One of the reasons for this is that state-of-the art theoretical approaches only provide incomplete descriptions of the electronic and magnetic states in the presence of strong electronic correlations.[3] But equally important is the scarcity of clean model systems (i.e., without chemical disorder or mixed-valence TMOs), which are crucial for comparing theoretical predictions with experimental data. The nickelate perovskite family RENiO$_3$ (RE = trivalent 4$f$ lanthanide and Y$^{3+}$) constitutes a unique, particularly well suited, and widely used system for this kind of investigations[3-7] because, in contrast with most TMO systems, a complete evolution from a paramagnetic metal to an antiferromagnetic insulator can be realized without doping.[8] LaNiO$_3$, the first member of the series, is paramagnetic, metallic and crystallizes in the rhombohedral $R\bar{3}c$ space group (SG).[9, 10] The remaining nickelates undergo spontaneous MITs at temperatures $T_{MIT}$ that increase as the size of the RE-cation decreases, (Figure 1).[11-16] Long-range Néel order of the Ni magnetic moments has been observed at temperatures $T_N \leq T_{MIT}$, which coincide with the electronic localization in the case of PrNiO$_3$ and NdNiO$_3$ (Figure 1).[10, 13, 17-22] RENiO$_3$ perovskites are thus intrisicaly close to the boundary between localized and itinerant behavior, marked by the $T_{MIT}$ vs RE radius line in Figure 1. This boundary can be crossed from both sides by changing either temperature, or the RE cation size (Figure 1), two clean, convenient variables for both, experimental and theoretical studies.

To date, most investigations on RENiO$_3$ nickelates have been focused on the mechanism at the origin of the MIT, which has been matter of an intense debate since its discovery in 1991.[11] The unusually high Ni oxidation state, formally Ni$^{3+}$ low spin (LS) $t_{2g}^6 e_g^1$ and potentially Jahn-Teller active, suggested an electronic instability as a possible origin of the transition. On the other hand, the implication of the lattice has been confirmed by the giant $^{16}O$-$^{18}O$ isotope effect on $T_{MIT}$[23] and by diffraction studies, which revealed a subtle symmetry breaking from orthorhombic *Pbnm* to monoclinic $P2_1/n$ below $T_{MIT}$ that splits the unique Ni site of the high-temperature metallic phase into two inequivalent Ni sites with slightly different average Ni-O distances.[15, 24-28] This breathing distortion, that contrasts with the expected deformation of the NiO$_6$ octahedra observed in other $e_g^1$ Jahn-Teller ions, has been interpreted as evidence of either a $2Ni^{3+} \rightarrow Ni^{3+\delta} + Ni^{3-\delta}$ charge disproportionation (CD),[24] or a Ni-O bond disproportionation with constant charge (2+) at the Ni sites and different ammounts of holes at the O sites.[29] Concerning the role of magnetism, the situation is less clear, in particular because $T_{MIT}$ and $T_N$ do not always coincide. However, there is increasing evidence suggesting that the presence of the magnetically ordered state just below $T_{MIT}$ may have an impact on the order of the transition.[30-32]

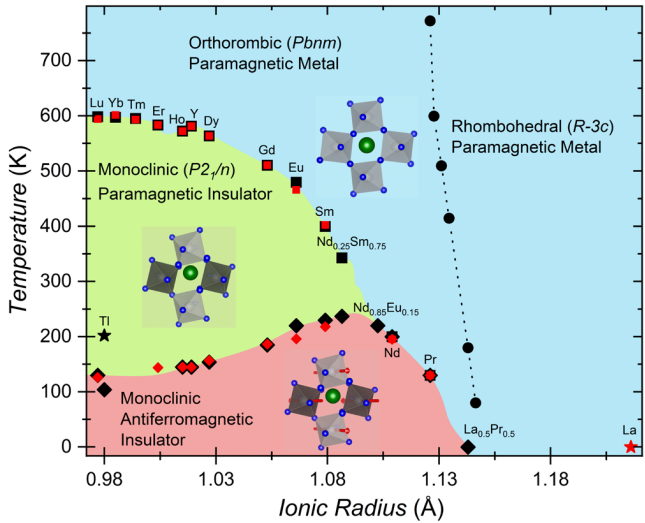

**Figure 1.** Phase diagram of bulk RENiO$_3$ perovskites. The red markers indicate the $T_{MIT}$ (squares) and $T_N$ (rhombuses) values obtained in this study. The black squares, rhombuses and circles are respectively the $T_N$, $T_{MIT}$ and *Pbnm* → $R\bar{3}c$ structural transition values previously reported in the literature for bulk polycrystalline samples.[10-22, 33] Red star: LaNiO$_3$ ($T_{MIT}$ = $T_N$ = 0 K). For nickelates with monoclinic $P2_1/n$ symmetry at base temperature, the ionic radii in the abscissa correspond with the values tabulated in Ref.[34] for trivalent $4f$ cations in 8-coordination. For rhombohedral LaNiO$_3$ with $R\bar{3}c$ symmetry at base temperature, we use instead the ionic radii for trivalent $4f$ cations in 9-coordination.

The simultaneous presence of lattice, electronic and magnetic instabilities offers a unique opportunity to investigate the role of the three most relevant degrees of freedom that control the properties of most TMOs. At the same time, it makes a full understanding of the MIT mechanism a formidable task that challenges the state-of-the art theoretical developments, such as Density-Functional[35] and Dynamical Mean-field Theory[36] methods. Ideally, the formalism and the values of the fundamental interactions used in the calculations could be improved by comparing theoretical predictions with experimental data, in particular with the phonon, magnon, and electronic dispersion curves along different reciprocal space directions. However, this information is very limited due to the scarcity of nickelate single crystals. Although some band structures have been reported for nickelate thin films,[37, 38] their properties differ in general from those of bulk samples due to their strong dependence on variables suchs as the film thickness or the lattice mismatch with the substrate. Moreover, thin films are not well suited for detailed diffraction studies, making difficult a detailed characterization of their crystal structures, which are needed as input in theroretical studies. A further limitation is that good quality epitaxial thin films have only been reported to date for the early nickelates (RE = La to Eu), and just for a few substrate orientations.[39, 40] Similar limitations also apply to magnon studies, where the available data are even scarcer, and only available for RE = Nd.[41, 42] Phonon dispersions have not yet been reported, as far as we know. There is thus an urgent need for sizable, good quality bulk single crystals of the full nickelate series.

The reason behind the reduced single crystal availability has been the difficult synthesis procedure for bulk RENiO$_3$ perovskites, which require highly oxidizing conditions in order to stabilize the unusually high Ni formal valence (3+). The synthesis of the (nearly complete) RENiO$_3$ family (RE = Y, La, Nd, Sm, Eu, Gd, Dy, Ho, Er, Tm, Yb, Lu) was first reported by Demazeau et al. in 1971,[19] who prepared them in polycrystalline form at 950°C and 60 kbar of hydrostatic pressure in a Belt apparatus by adding KClO$_3$ as an oxygen source. The nickelates with RE = Ce, Pr and Tb could not be stabilized under such elevated pressures due to the amphoteric character of these cations, which can display both, +3 or +4 oxidation states. Twenty years later, an alternative approach based in the use of much lower oxygen gas pressures (up to 200 bar) and high temperatures (1000°C) allowed Lacorre et al. to obtain PrNiO$_3$ for the first time.[11] NdNiO$_3$, SmNiO$_3$, EuNiO$_3$ and GdNiO$_3$[11, 12, 43-46] could be also prepared in this way at slightly different temperatures using oxygen gas pressures up to 400 bar. This pressure was not enough to stabilize the nickelates with smaller RE cations (Dy to Lu), but Alonso et al.[13, 14, 47] proved the possibility of synthesizing them at hydrostatic pressures (20 kbar) three times lower than those used by Demazeau, although in very small amounts (< 1 g). The syntheses were conducted in a piston-cylinder press at 900°C with KClO$_4$ as an oxygen source.

In line with the difficulties of preparing polycrystalline samples, obtaining good quality RENiO$_3$ single crystals has been to date extremely challenging. Attempts to growth PrNiO$_3$ and NdNiO$_3$ in different fluxes (KCl / NaCl / KClO$_4$ / NaClO$_4$ / NaOH) under high hydrostatic pressures (up to 40-45 kbar) and temperatures in the 1400-1500°C range have been reported in the past.[48-50] Crystals up to 500 μm (Pr) and 100 μm (Nd) could be obtained, but they suffered from twinning due to the small deviations of their crystal structure from that of the cubic perovskite.[51] Recently, large LaNiO$_3$[52, 53] and PrNiO$_3$[54] single crystals were grown by the travelling solvent floating zone (TSFZ) technique under moderate oxygen pressures (up to 150 bars). However, the oxygen stoichiometry was found to be inhomogeneous in both axial and radial growth directions[55]. This had a huge impact on $T_{MIT}$ and the magnetic behavior, very different from those reported for stoichiometric ceramic samples.

Here we report the first successful growth of RENiO$_3$ single crystals up to ~75 μm in size for RE = Nd, Sm, Gd, Dy, Y, Ho, Er and Lu with prismatic shapes, flat facets, and identical metal-insulator and antiferromagnetic transition temperatures to those of the polycrystalline samples of the same composition (Figure 1). We used an original method based in the use of moderate oxygen gas pressures (2000 bar), solvothermal growth in a temperature gradient, and highly reactive eutectic salt mixtures as fluxes that can be used to growth RENiO$_3$ nickelate crystals covering the full $4f$ series. The availability of such crystals will grant access to properties never reported to date for bulk samples. This should provide crucial insights to theoretical studies, contribute to a better understanding of the complex physicochemistry of correlated transition metal oxides, and pave the way to a targeted design of novel materials with strategically relevant magnetic and electronic properties.

RESULTS

**High pressure crystal growth.** The growth under such extreme conditions was carried out in a unique, non-comercial high oxygen gas pressure apparatus recently installed at the Paul Scherrer Institute (PSI) in Villigen, Switzerland.[56] As shown in Figure 2, the technical challenge of combining oxygen gas pressures up to 2000 bar with temperatures up to 1000°C is solved by minimizing the oxygen volume, whose pressure is continuoulsly equilibrated by an inert gas (Ar) counter pressure. Another distinctive characteristic of the device is the large size of the main vessel (~ 50 cm³), which allows solid state synthesis of ceramics, as well as solvothermal growth of single crystals, in large amounts.

For each nickelate, about 5g of nano-crystalline precursors were prepared by mixing stoichiometric amounts of rare earth RE$_2$O$_3$ and NiO (SI1, Materials and Methods), previously dried at 900 °C and at 400 °C, respectively. The mixed dehydrated oxides were dissolved in 75 ml of concentrated 65-75% HNO$_3$ (aq.). Water and nitric acid were removed by continuous stirring and heating the solution at 350 °C. The solid residue was heated in a muffle furnace in air at 300 °C for three hours, and the obtained black powder was further annealed at 650 °C for 24 h in an alumina crucible under a 200 ml / min oxygen flow.

The resulting materials were mixed with a LiCl/KCl flux in a helium-filled glove box in the proportion 1:5 and placed into a semi-closed alumina reactor (internal diameter 11 mm and length 130 mm). For the flux, a eutectic mixture LiCl/KCl (4/6) was chosen for its relatively low melting point (~450 °C), inert reactivity towards the RENiO$_3$ precursors and easy workup after the reaction (dissolves in water). An additional advantage is that it can play the role of an electrolyte, and thanks to that, the oxidation process can be boosted, increasing the kinetics and decreasing the thermodynamic parameters. The reactor was introduced into the high-pressure furnace (see scheme and picture in Figure 2) and heated up to 850°C under 2000 bar oxygen gas pressure for 12 h. The temperature gradient between the bottom and top of the semi-closed reactor was ~ 100 °C. Under these conditions, phase pure RENiO$_3$ microcrystals with RE = Nd, Sm, Gd, Dy, Y, Ho and Er could be obtained. To separate them from the flux, the final reaction products were washed with distilled water (to dissolve the LiCl/KCl eutectic), ethanol, and acetone, after which they were dried in air.

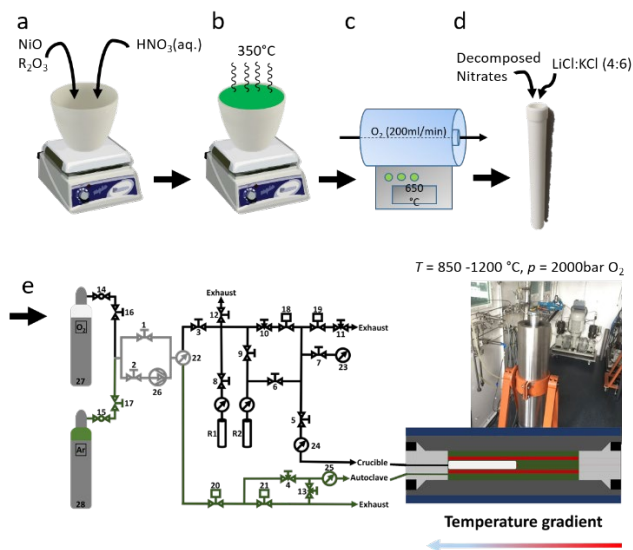

**Figure 2.** High-pressure single crystal growth process. (a) Dissolution of the starting oxides in nitric acid (65%, aq.); (b) evaporation of water and decomposition of NO$_2$/NO$_3$; (c) oxidation under O$_2$ flow; (d) mixing of precursors and solvent in a high pressure crucible; (e) schematic representation of the high-pressure, high-temperature furnace with the temperature gradient and Ar/O gas control setup (hand operated needle valves (1,-3, 5-9, 12, 16, 17); hand operated micro metering valves (4, 10, 11, 13); air operated valves (18-21); manometers (22-25); gas compressor (26); oxygen gas cylinder (27) with valve (14); argon gas cylinder (28) with valve (15); pressurised gas reservoirs (R1, R2)), together with a picture of the reactor.

The nickelates with smaller rare earths (RE = Tm, Yb) could not be prepared in pure form using this procedure. As shown in Figure S1, large amounts of unreacted RE$_2$O$_3$, NiO, REClO and/or mixed Li/Ni oxides were found to coexist with the RNiO$_3$ phases, which totalized ~ 5% of the final product' weight. Interestingly, the lattice parameters and $T_{MIT}$ values of both, TmNiO$_3$ and YbNiO$_3$ were nearly identical to those reported in the literature (Figure S1). This suggested a kinetic origin for the incomplete RENiO$_3$ formation reaction, further confirmed by the increase of the TmNiO$_3$ weight fraction to ~25% after a second annealing (Figure S1b). In an attempt to speed up the process, a synthesis was performed for LuNiO$_3$ using a more oxidizing LiClO$_4$/KClO$_4$ (4/6) flux. Under the action of temperature and pressure the perchlorates decomposed, yielding the original LiCl/KCl (4/6) flux plus extra oxygen in the form of highly-reactive free radicals. After removing the flux, the final product contained phase-pure LuNiO$_3$ microcrystals (~90% in weight), together with ~10% ) of unreacted Lu$_2$O$_3$. Crystalline NiO and/or mixed Li/Ni oxides could not be observed, but they were probably present in nanocrystalline or amorphous form.

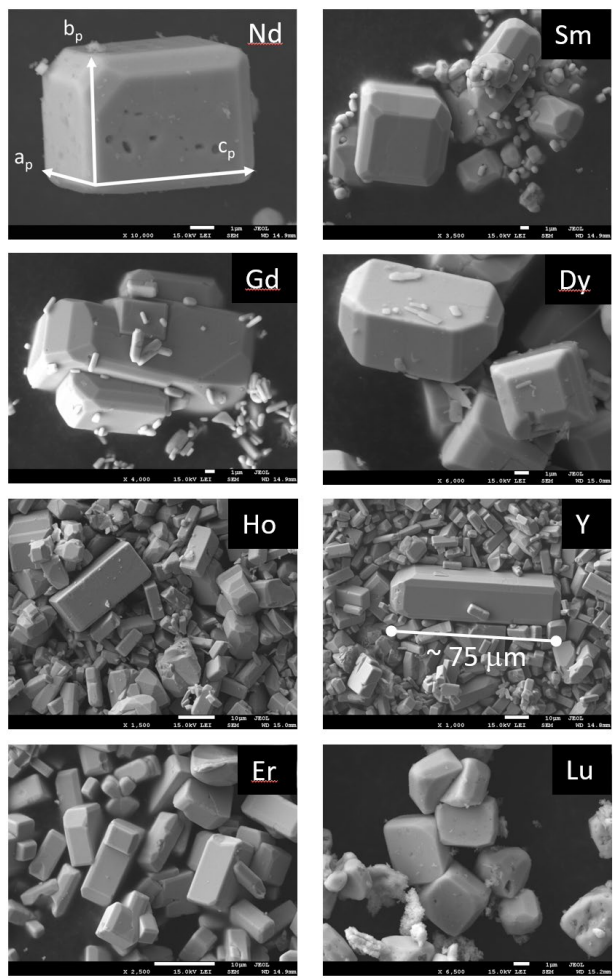

**Figure 3** SEM images of the RENiO$_3$ microcrystals with RE = Nd. Sm, Gd, Dy, Y, Ho, Er and Lu. The most common facet orientation is illustrated for a NdNiO$_3$ crystal, where $a_p$, $b_p$ and $c_p$ denote the pseudocubic perovskite lattice parameters.

The single crystal nickelate set was completed with polycrystalline LaNiO$_3$, PrNiO$_3$, and EuNiO$_3$ samples. The PrNiO$_3$ and EuNiO$_3$ samples were the same used in our previous works,[18, 32] where the synthesis route was already reported. LaNiO$_3$ was obtained by solid-state reaction at 950 °C under 2000 bar oxygen pressure starting from the oxide precursors.

The crystal habit, size and atomic composition of the crystals were characterized using Field-Emission Scanning Electron Microscopy (FE-SEM) coupled with Energy-Dispersive X-ray Spectroscopy (EDX). The phase purity and crystal structure of all samples were investigated by laboratory powder X-ray diffraction, plus single-crystal X-ray diffraction in the case of the larger crystals. Differential Scanning Calorimetry (DSC), SQUID magnetometry and Heat Capacity (C$_p$) were employed to characterize their physical properties and to determine their metal-insulator and antiferromagnetic order temperatures. Further details are provided in the Supporting Information.

**Crystal habit, size and composition.** The materials obtained after the flux removal had the appearance of fine powders, but examination under the optical microscope revealed tiny, well-separated microcrystals with sizes up to ~75 μm. In spite of the increasing difficulty of stabilizing the RENiO$_3$ phases for the nickelates with smaller RE cations, the crystals were in general larger and more regular for YNiO$_3$, HoNiO$_3$ and ErNiO$_3$ (Figure S2). These observations were confirmed by the SEM images (Figure 3), which revealed the presence of crystals with well-developed flat facets, most of them with a regular, truncated prism shape. This crystal habit was very different from that of the REClO phases present in some of the preparations and easily identifiable owing to their characteristic diamond-like shape (Figure S3). The chemical composition of the crystals (REClO and RENiO$_3$) was checked by EDX, which confirmed the nominal stoichiometry (Figures S3 and S4).

The facet orientation of the RENiO$_3$ phases was checked for some of the largest exemplars using single-crystal X-ray diffraction. In most of them, the largest crystal facets were perpendicular to the pseudocubic perovskite crystal axes $a_p$ and $b_p$, with the longest edge parallel to the $c = 2a_p$ axis of the orthorhombic *Pbnm* unit cell, as shown in Figure 3. The $a = a_p − b_p$ and $b = a_p + b_p$ *Pbnm* axes, parallel to the in-plane diagonals of the psudocubic perovskite cell, were thus perpendicular to the truncated facets. Well-developed facets were in general not observed in crystals with sizes smaller than 0.5 μm. Recrystallization attempts aimed at increasing the nickelate crystal size were conducted in the case of SmNiO$_3$ by repeating the growth procedure, but they were unsuccessful. Instead, larger amounts of SmClO were observed, with crystal sizes notoriously larger than in the original preparation (Figure S5). We do not exclude that larger nickelate crystals can be prepared in the future, but this will require additional experimental work.

**Crystal structure.** As mentioned in the introduction, all nickelates in polycrystalline form, with the exception of LaNiO$_3$, feature sharp MITs with pronounced anomalies in the lattice parameters. However, the subtle *Pbnm* → *P2$_1$/n* symmetry-breaking below $T_{MIT}$ and the associated splitting of the Ni sites has only be reported for some members of the family (RE = Pr[25], Nd[28], Sm[27], Dy-Lu[15, 16, 26]). The reason behind this is the smallness of the monoclinic angle *β*, that at RT takes values between ~90.16 (Lu)[47] and ~90.04° (Sm)[27] - NdNiO$_3$ and PrNiO$_3$ are metallic and orthorhombic at RT, and is thus difficult to detect with standard laboratory X-ray diffraction techniques. Only high resolution powder diffraction (neutron or X-ray synchrotron) was able to reveal the tiny monoclinic distortion and the associated splitting of the Ni sites in the insulating state, whose crystal structure was described as orthorhombic *Pbnm* in early studies at both sides of the MIT. Although the existence of a monoclinic distortion below $T_{MIT}$ has not yet been reported for EuNiO$_3$ and GdNiO$_3$, evidence supporting a direct link between the MIT and the existence of two distinct Ni sites in the insulating state for the whole RENiO$_3$ family has been provided by other techniques, such as X-ray absorption[57] and Mössbauer spectroscopies.[16, 26] emerges as the main distinctive characteristic of the insulating phase, as well a necessary condition for the electronic localization below $T_{MIT}$.

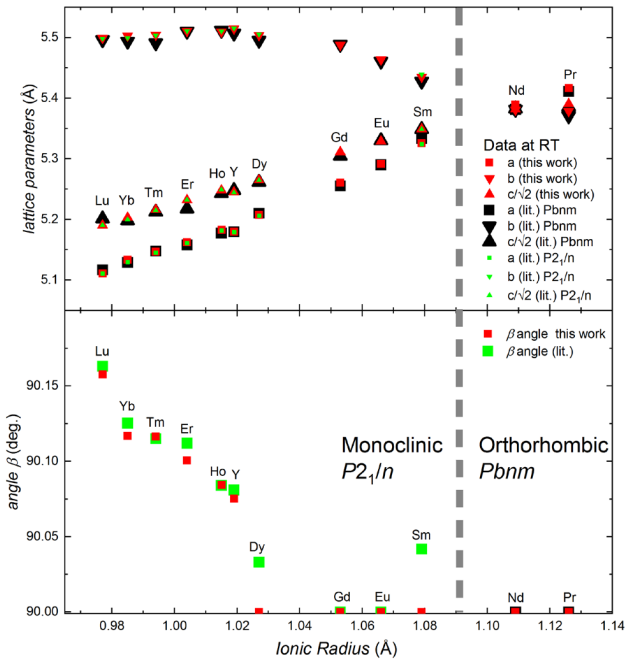

The results of the structural characterization of our nickelate samples fully agrees with this picture for both ceramics (RE = La, Pr and Eu) and single crystals. As shown in Figure S6, the laboratory powder X-ray diffraction patterns of LaNiO$_3$, PrNiO$_3$ and NdNiO$_3$ could be indexed on the basis of a rhombohedral $R\bar{3}c$ (La) or an orthorhombic $Pbnm$ lattice (Pr and Nd), in agreement with previous studies. For SmNiO$_3$ and DyNiO$_3$, reported to be insulating and monoclinic at RT the tiny angle $\beta$ reported in previous works (~90.04° [27] and ~90.03° [26], respectively) could not be resolved by our laboratory equipment. This was also the case for EuNiO$_3$ and GdNiO$_3$, where the existence of a monoclinic distortion has not been reported to date. A clear splitting of the $(hkl)$ and $(-hkl)$ reflections consistent with a monoclinic angle $\beta \neq 90°$ was, in contrast, clearly observed for YNiO$_3$, HoNiO$_3$, ErNiO$_3$, TmNiO$_3$, YbNiO$_3$ and LuNiO$_3$, which were reported to be insulating and monoclinic at RT with substantially larger $\beta$ angles. The Rietveld fits, shown in Figures S1 and S6, were thus conducted with the SG $Pbnm$ for the nickelates with with RE = Pr, Nd, Sm, Eu, Gd and Dy, and with $P2_1/n$ for those with RE = Y, Ho, Er, Tm, Yb and Lu. However, the observation of MITs with transition temperatures nearly identical to those reported in the literature for all nickelates with R ≠ La (see Figure 1 and the section "metal insulator transitions") strongly suggest that SmNiO$_3$, EuNiO$_3$, GdNiO$_3$ and DyNiO$_3$ are actually monoclinic at RT.

**Figure 4.** RT lattice parameters $a$, $b$ and $c$ and $\beta$ of RENiO$_3$ perovskites as a function of the RE ionic radius. The red markers are the values obtained in this study. The remaining markers are data reported in the literature for ceramic samples using the $Pbnm$ (black)[19] and $P2_1/n$ (green)[26, 28, 47] groups.

This scenario is also supported by most theoretical studies,[4, 6, 7, 58, 59] where the presence of small and large NiO$_6$ octahedra

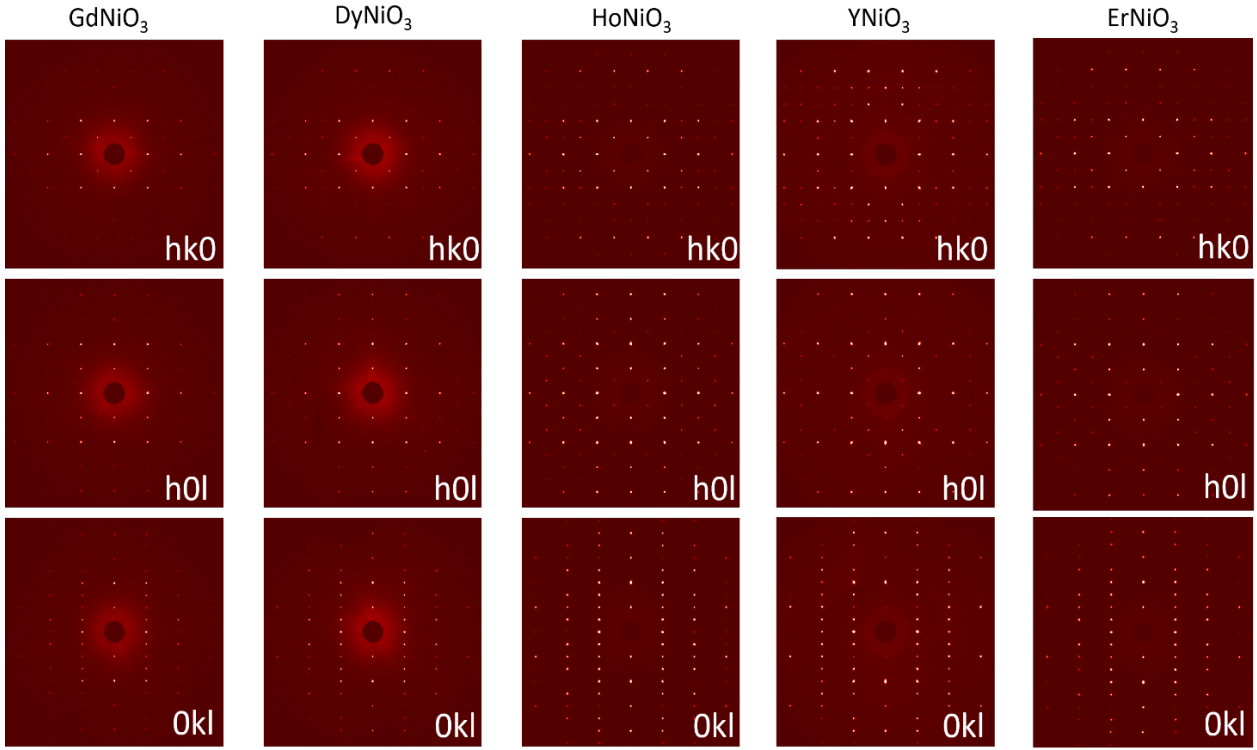

**Figure 5.** Reciprocal (hk0), (h0l) and (0kl) lattice planes measured by laboratory X-ray diffraction (Mo $K\alpha$) at RT for nickelate single crystals with RE = Gd, Dy, Ho, Y and Er, illustrating the agreement between the observed systematic absences and those of $Pbnm$.

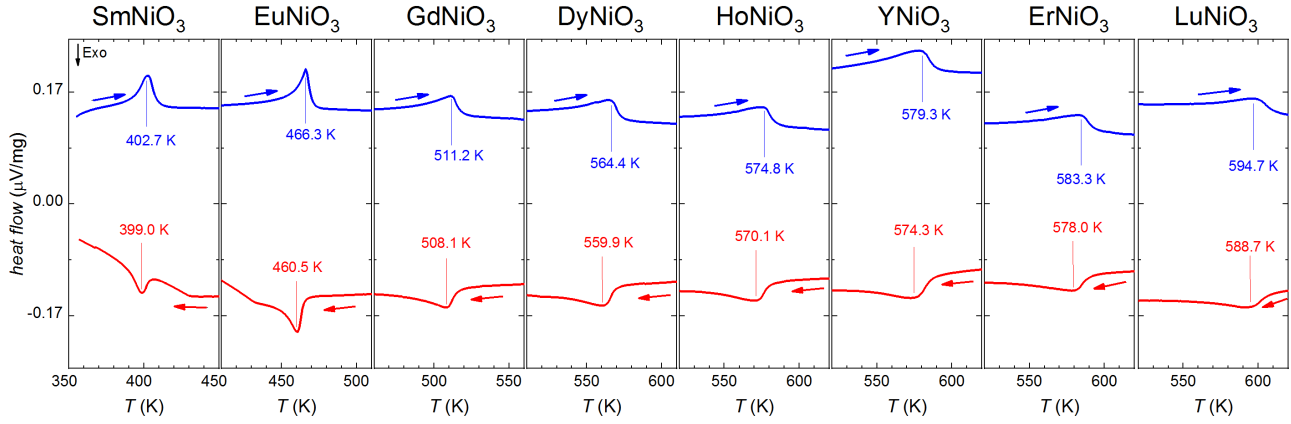

**Figure 6.** DSC curves obtained for the RENiO$_3$ single crystals (RE = Sm, Gd, Dy, Y, Ho, Er and Lu) and polycrystalline EuNiO$_3$ in a heating/cooling cycle.

The refined structural parameters, interatomic distances, and angles, very similar those reported in previous studies, are summarized in Table S1. The excellent agreement is illustrated in Figure 4, where the *a*, *b*, *c* l and *β* lattice parameters of our samples (red symbols) are compared with those of refinements performed using the *Pbnm* and *P2$_1$/n* SGs (black and green symbols, respectively).

The powder diffraction data were complemented with laboratory single-crystal X-ray diffraction data sets for the best, untwined exemplars (see the SI for further details on data collection and analysis). The highly single crystalline quality of representative GdNiO$_3$, DyNiO$_3$, YNiO$_3$, HoNiO$_3$ and ErNiO$_3$ specimens is illustrated in Figure 5, showing selected high-symmetry reciprocal lattice planes reconstructed from the RT data sets. The lattice metrics were compatible with the *mmm* Laue class and the systematic absences consistent with those of *Pbnm* for the five nickelates, including those whose monoclinic symmetry at RT was confirmed by PXRD. This was an expected result given the smallness of the monoclinic distortion and the superior 2*θ* resolution of powder diffractometers, which are better suited than single crystal diffractometers for the determination of the lattice geometry. Their crystal structures were thus refined using *Pbnm*. The obtained structural data, summarized in the deposited CIF files, are consistent with those obtained from the powder diffraction data (see the SI for further details and for the Cambridge Crystallographic Data Centre deposition numbers). The crystal structures of the five nickelates together with the associated atomic displacement ellipsoids (50 % probability) are shown Figure S7.

For NdNiO$_3$, SmNiO$_3$ and LuNiO$_3$, crystal sorting, mounting, data collection and analysis were more challenging. This was due to their smaller crystal size (Figure 3), the presence of impurities (RE = Lu) and in the case of RE = Nd amd Sm, to the frequent presence of twins. An in depth single-crystal diffraction investigation of these three nickelates will require additional experimental work and will be presented separately.

**Metal-to-insulator transitions.** Besides diffraction data, a central parameter for assessing the match between the properties of our crystals and those of previously reported ceramic samples is the value of the metal-insulator temperature $T_{MIT}$. For the nickelates with the smallest rare earth ions (RE = Sm, Eu, Gd, Dy, Y, Ho, Eu and Lu), where the metal to insulator transition was reported to occur above 400K, $T_{MIT}$ was determined using DSC. This technique, also employed in refs.[12, 13, 15, 23], was used instead of resistivity measurements for two reasons: the $T_{MIT}$ values were above the operation limit of our equipment (a PPMS 9T with $T_{max}$ = 400 K), and the largest RENiO$_3$ single crystals, too small to be contacted manually, could not be pressed or sintered to form dense pellets. DSC can easily access the required temperature range (400K < T < 600K), and provides a direct measurement of the enthalpy (*H*) change at the metal-to-insulator transition. This is related to the heat capacity at constant pressure ($C_p$) through the relationship $C_P = \frac{1}{m}\left(\frac{\partial H}{\partial T}\right)_P = \frac{1}{m}\frac{\Delta P}{\beta}$, where *m* is the sample mass, *β* is the heating rate $\partial T/\partial t$ during the DSC measuremenst, and *ΔP* is the absolute value of the heat flow through the sample (i.e., the absolute value of the DSC signal). Above 400K, the electronic and magnetic contributions to $C_P$ are in general very small, and the largest contribution comes from the lattice. Since the MIT in nickelates is also a first-order structural transition involving a SG change and pronounced discontinuities in the lattice parameters, it gives rise to an anomaly in the $C_P$ that can be easiy detected in the DSC signal.

Figure 6 shows the DSC signals measured by heating (in blue) and cooling (in red) of these nickelates, confirming the presence of endothermic peaks in the heating curves, with maxima between 402.7 K (Sm) and 594.7 K (Lu). These values, plotted as red symbols in Figure 1 and summarized in Table S2, are nearly identical to the metal-insulator temperatures reported in the literature and obtained using different techniques (black symbols in Figure 1). The exothermic peaks observed in the cooling curves display maxima at temperatures a few degrees lower, which indicate the presence of hysteresis. This behavior, characteristic of first order transitions, has been observed at $T_{MIT}$ for all nickelates, further supporting the assignment of the DSC peaks to the MIT's.

For PrNiO$_3$ (already reported in our previous study[32]) and NdNiO$_3$, where the electronic and magnetic transitions coincide, $T_{MIT} = T_N$ was determined from bulk magnetic susceptibility $\chi = M/\mu_o H$ measurements (Figure 7). For both nickelates, the $\chi(T)$ curves are dominated by the paramagnetic contribution of Pr$^{3+}$ and Nd$^{3+}$, with nominal $\mu_{eff}$ = 3.58 and 3.62 $\mu_B$ values, respectively (Table S3). However, tiny anomalies at 130.1 K (Pr) and 196.2 K (Nd), better appreciated in the $1/\chi$ curves, are clearly observed. As shown in Figure 1, these values agree very well with those previously reported from neutron diffraction.

**Magnetic transitions.** Complementary to $T_{MIT}$, the antiferromagnetic ordering temperatures of the Ni and RE

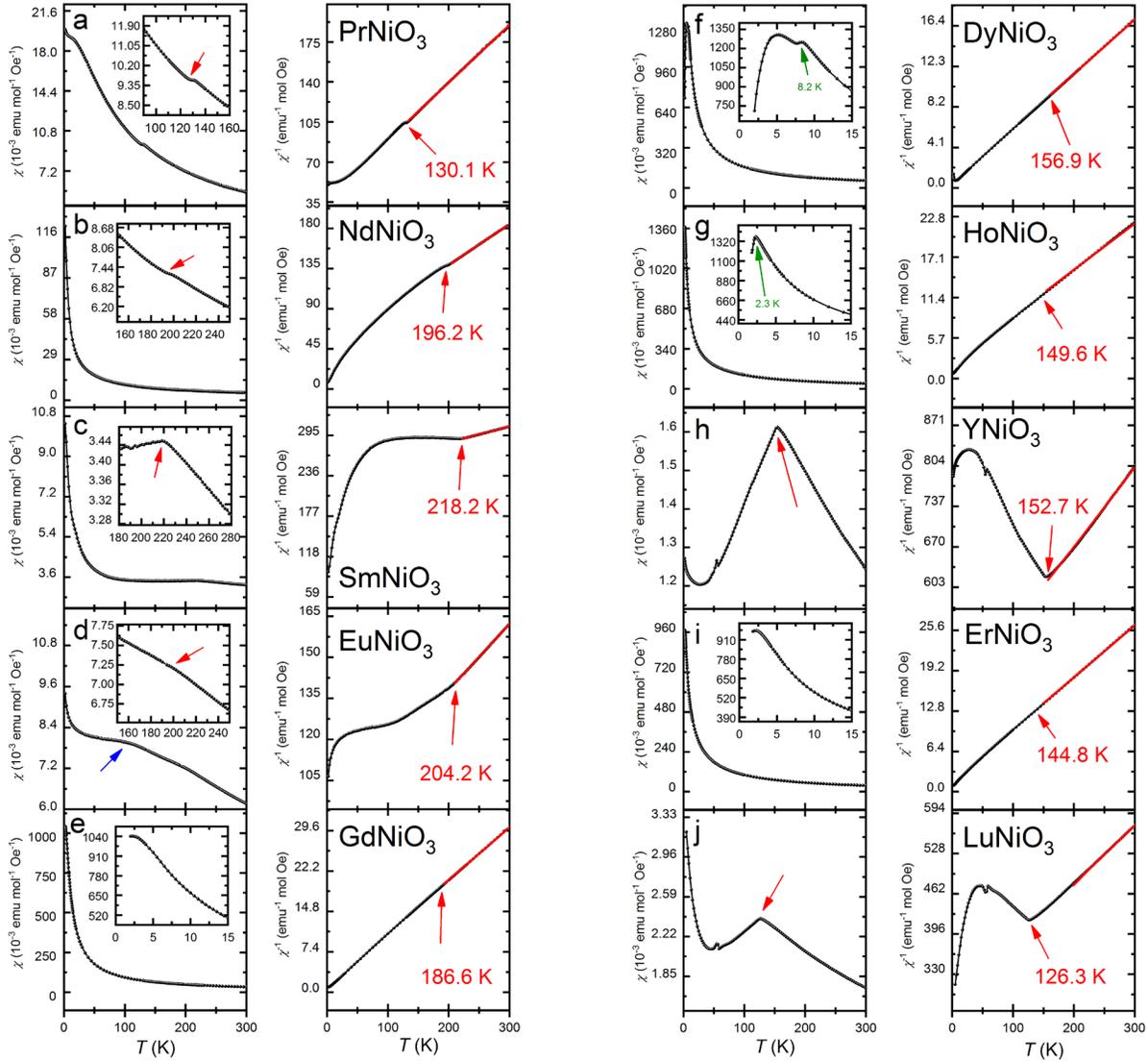

**Figure 7.** Temperature dependence of the magnetic susceptibility and its inverse, as obtained for our RENiO$_3$ single crystals (RE = Nd, Sm, Gd, Dy, Ho, Y, Er, Lu) and ceramics (RE = Pr, Eu). The red arrows indicate the antiferromagnetic ordering temperature $T_N$ of the Ni sublattice and the green arrows in (f) and (g) those of the Dy and Ho sublattices. The blue arrow indicates the inflexion point associated to the van Vleck paramagnetic signal in EuNiO$_3$. The Red lines indicate the temperature range used for the Curie-Weiss fits. The Pr data are from Ref.[32]

sublattices offer an additional way to assess the properties of our crystals. We thus performed additional magnetic susceptibility measurements for the nickelates with RE ≠ Pr, Nd, where $T_N$ and $T_{MIT}$ are well separated and $T_N$ is expected to be below RT. As shown in Figure 7, clear anomalies in the $\chi(T)$ and $1/\chi(T)$ curves could be observed for SmNiO$_3$ (218.2 K), EuNiO$_3$ (204.2 K), YNiO$_3$ (152.7 K) and LuNiO$_3$ (126.3 K), very similar to those reported in previous studies on powder samples (Figure 1). For the nickelates with RE = Gd, Dy, Ho and Er, no anomaly was evident in the $\chi(T)$ curves due to the huge paramagnetic contribution of the RE$^{3+}$ magnetic moments, which mask the weak signal of the Ni sublattice. The $T_N$ values were thus determined from heat capacity ($C_p$) measurements as a function of temperature, as shown in Figure 8. Tiny peaks at $T$ = 186.6 K, 156.9 K and 149.6 K are clearly observed for GdNiO$_3$, DyNiO$_3$ and HoNiO$_3$, respectively, and are very similar to the $T_N$ values reported in refs.[13, 20] (Figure 1). For ErNiO$_3$, where an experimental $T_N$ value has not been reported so far, a similar anomaly is observed at 144.8 K, very close to the extrapolated $T_N$ value reported in ref.[8].

For DyNiO$_3$ and HoNiO$_3$, additional anomalies with sharp maxima at $T \sim 8.2$ K and ~2.3 K were observed in the $\chi(T)$ curves. These values are very similar to the long-range ordering temperatures of the Dy$^{3+}$ and Ho$^{3+}$ magnetic moments determined by neutron powder diffraction[20, 26], further confirming the excellent agreement between the physical properties of our samples and those of previously reported powder samples.

**Effective paramagnetic moments and Curie-Weiss temperatures.** Although an in-depth characterization of the RE and Ni magnetism in RENiO$_3$ perovskites is out of the scope of this work, the analysis of the inverse magnetic susceptibility $1/\chi$ in the paramagnetic region ($T > T_N$, see Figure 7 and the SI for details) provides a first hint about the sign and size of the dominant magnetic interactions and the effective paramagnetic moments $\mu_{eff}$ along the series. This information, never reported to date for the full nickelate family, is complementary to that obtained from neutron diffraction, which probes the moment values in the magnetically ordered state.

For the nickelates without magnetic RE cations (RE = Y, Lu), $1/\chi$ was modelled with a Curie-Weiss law $\frac{1}{\chi} = \frac{T-\theta}{C_{Ni}}$, where $\theta$ is the Curie-Weiss temperature and $C_{Ni}$ is the Curie constant, related to the spin-only Ni effective paramagnetic moment $\mu_{eff}$ through the expression $\mu_{eff} = \sqrt{\frac{3k_B C_{Ni}}{N_A \mu_B^2}} \sim 2.828\sqrt{C_{Ni}}$. Here, $k_B$ is the Boltzmann constant, $N_A$ is the Avogadro number, and $\mu_B$ is the Bohr magneton. The obtained $\mu_{eff}$ values where then compared with those reported in neutron diffraction studies. In the case of YNiO$_3$, the paramagnetic moments of the Ni at the "large" and "small" octahedral sites reported in ref.[24] are 1.4 $\mu_B$ and 0.6 $\mu_B$. These values are respectively smaller and larger than those of Ni$^{2+}$ ($t_{2g}^6 e_g^2$, 2 $\mu_B$) and Ni$^{4+}$ ($t_{2g}^6 e_g^0$, 0 $\mu_B$), which, in a simple ionic picture, was interpreted as a signature of partial Ni charge disproportionation. From the Curie fits of our YNiO$_3$ and LuNiO$_3$ samples (Figure 7) we obtain $\mu_{eff}$ = 2.62 $\mu_B$ (Y) and 2.96 $\mu_B$ (Lu). For YNiO$_3$, this value is very close to the one expected for the degree of disproportionation determined from neutron diffraction ($\mu_{eff}$ = 2.51 $\mu_B$), and much larger than the effective paramagnetic moment of non-disproportionated Ni$^{3+}$ LS ($\mu_{eff}$ = 1.73 $\mu_B$). At the same level of approximation, this suggest that the moments (and the charges) of both Ni sites remain different in the paramagnetic region close of $T_N$, with a degree of disproportionation similar to that observed in the antiferromagnetic state. Another interesting result is the high value of the Curie-Weiss temperatures, $\theta$ = -245 K (Y) and -500 K (Lu), which indicate the presence of large, predominantly antiferromagnetic interactions in the Ni sublattice. The Ramirez ratios, $|\theta|/T_N$ = 1.60 (Y) and 3.96 (Lu), also suggest the existence of weak magnetic frustration, which increases as the RE ionic radii decrease.

For the nickelates with RE = Pr, Nd, Gd, Dy, Ho and Er, the $\chi(T)$ curves are dominated by the paramagnetic signal of the RE$^{3+}$ cations. Given the impossibility of extracting the Ni contribution independently, we carried out the Curie-Weiss fits by assuming the simultaneous presence of Ni$^{3+}$ LS with spin-only moment ($\mu_{eff}$ = 1.73 $\mu_B$). $1/\chi$ was thus modelled with a Curie-Weiss law $\frac{1}{\chi} = \frac{T-\theta}{C}$, where $C = C_{RE} + C_{Ni}$ is the fitted Curie constant, $\theta$ is the Curie-Weiss temperature, $C_{Ni}$ = 0.375 emu K mol$^{-1}$ is the Ni contribution, and $C_{RE}$ the contribution of the RE cation.

The $\mu_{eff}$ values obtained in this way are 3.583 (Pr), 3.798 (Nd), 8.938 (Gd), 11.94 (Dy), 10.95 (Ho), and 9.704 $\mu_B$ (Er), which, as expected, are slightly larger than the nominal free-ion values (Table S3). The Curie-Weiss temperatures, $\theta$, are -76.2 (Pr), -102.8 (Nd), -12.8 (Gd), -10.1 (Dy), -35.0 (Ho), and -20.2 K (Er). As in the case of the Ni-Ni coupling, the dominant magnetic interactions when both Ni and a magnetic RE cation are present are also antiferromagnetic, but with significantly smaller values.

For SmNiO$_3$ and EuNiO$_3$, the smaller energy separation between the ground and the first excited $J$-multiplets of the RE$^{3+}$ cations results in some admixture of the excited state in the ground state, giving rise to an additional van Vleck-type contribution to the paramagnetic response. The Curie-Weiss law is no longer obeyed in the full paramagnetic state, and if used at sufficiently high temperatures, leads in general to $\mu_{eff}$ values larger than the single-ion values and high Curie-Weiss temperatures, suggesting the presence unrealitistically large exchange interactions.[60] For Sm$^{3+}$, with an odd number of 4$f$ electrons, the lowest-laying $J$ multiplets ($^6H_{5/2}$ and $^6H_{7/2}$) are split by the CEF interaction into three and four Kramers doublets, respectively. The energy separation between the lowest-lying doublet of the ground state and that of the first excited multiplet, as determined from inelastic neutron scattering, is 132 meV (~ 1530 K).[33]

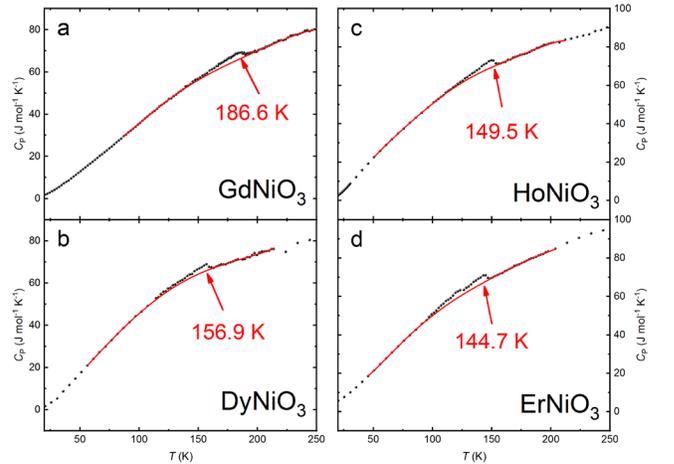

**Figure 8.** Temperature dependence of the heat capacity for the GdNiO$_3$, DyNiO$_3$, HoNiO$_3$ and ErNiO$_3$ single crystals. The red lines represent the phonon contribution estimates as described in the Methods section (SI).

The doublets of $^6H_{7/2}$ are thus expected to be partially populated below $T_N$ = 218.2 K, the temperature above which both Ni$^{3+}$ and Sm$^{3+}$ are paramagnetic. This is confirmed by the results of the Curie-Weiss fits, which lead to $\mu_{eff}$ = 5.63 $\mu_B$, a value that is much larger than the free ion value (0.85 $\mu_B$), and a very large $\theta$ = -1035.9 K. In the case of Eu$^{3+}$, with an even number of 4f electrons, the lowest-lying J multiplets are the singlet $^7F_0$ and triplet $^7F_1$ states. The ground state is not degenerate and the energy gap with the lowest-lying level of $^7F_1$ is only 34.5 meV (~ 400 K).[61] The admixture between ground and excited states is thus expected to be much more important than in the case of $Sm^{3+}$. This is supported by the shape of the $\chi(T)$ curve, where the characteristic, temperature-independent van Vleck contribution is clearly observed below ~100K (blue arrow in the EuNiO₃ panel of Figure 7). A small paramagnetic contribution is also present at lower temperatures, which we tentatively attribute to an unidentified paramagnetic impurity. As in the case of $Sm^{3+}$, a Curie-Weiss fit between $T_N$ and 300K led to $\mu_{eff}$ = 5.46 $\mu_B$, which is much higher than the free ion value (0 $\mu_B$), and a large $\theta$ = -371 K.

**Entropy loss associated with the antiferromagnetic order of the Ni magnetic moments.** A further parameter providing information about the Ni magnetic moment is the the entropy loss $\Delta S$ associated with the Ni cooperative order below $T_N$. For the nickelates with RE = Gd, Dy, Ho and Er, $\Delta S$ was estimated from the excess $C_p$ after subtraction of the phonon contribution (red curves in Fig. 8, see the S1 Methods section in the SI for details). The obtained entropy changes $\Delta S$ were 580, 566, 594 and 1052 mJ mol$^{-1}$ K$^{-1}$ for RE = Gd, Dy, Ho and Er, respectively. These values were then compared with the theoretical value expected from the ordering of $Ni^{3+}$ magnetic moments with $S$ = ½ ($\Delta S = R \ln 2$ = 0.69 $R$), where $R$ is the ideal gas constant. For our nickelates we obtain $\Delta S$ ~ 0.070 $R$ (Gd), 0.068 $R$ (Dy), 0.071 $R$ (Ho) and 0.127 $R$(Er). These values differ slightly from those reported by Alonso et al.[13] (~ 0.022 $R$ and ~ 0.090 $R$ for RE = Gd and Dy, respectively), although the discrepancy might be due to the smallness of the anomaly and the details of the phonon background subtraction. In any case, both our $\Delta S$ values and those of ref.[13] are much lower than the theoretical value of 0.69 $R$. Similar observations have been reported for SmNiO₃ ($\Delta S$ = 0.03 $R$),[62] and even for NdNiO₃ ($\Delta S$ = 0.12 $R$)[63] and PrNiO₃ ($\Delta S$ = 0.16 $R$),[32] where the magnetic, electronic and structural transitions coincide and extra contributions to $\Delta S$ would be expected. Although the entropy jump is indeed larger for these two nickelates, it is still far below the purely magnetic theoretical value. This suggests the existence of "lost" magnetic entropy above $T_N$ for the whole RENiO₃ family, most probably in the form of magnetic correlations that, in the case of PrNiO₃ and NdNiO₃, could survive even in the metallic phase above $T_N = T_{MIT}$.

CONCLUSIONS

To summarize, we have reported the first successful growth of phase-pure RENiO₃ single crystals with RE = Nd, Sm, Gd, Dy, Y, Ho, Er and Lu. The crystals have well-defined facets, regular, slightly truncated prismatic habit and sizes up to ~75 μm. We used an original method involving the use of a highly reactive molten salts flux, a high temperature gradient and oxygen gas pressures up to 2000 bar, delivered by an unique, non-commercial apparatus recently installed at the PSI. This technically challenging route involves the use of oxygen gas pressure five times larger than those employed in previous studies (P$_{max}$ ~400 bar), but ten times lower that the 20 kbar required to synthesize ceramics with RE = Dy to Lu using hydrostatic pressure. Moreover, the large size of the reactor allows the preparation of material amounts up to 10 times larger than in the case of hydrostatic pressure. The availability of these samples enabled the first systematic investigation of RENiO₃ bulk magnetic properties. This provided novel information on the evolution of the dominant magnetic interactions and the magnetic correlations in the paramagnetic phase along the series, and the first experimental determination of $T_N$ for ErNiO₃.

The nickelate single crystals reported in this study are unique in many senses. On one side, they are the first-ever RENiO₃ perovskites with RE = Sm, Gd, Dy, Y, Ho, Er and Lu grown in single crystalline form. On the other, their metal-to-insulator and antiferromagnetic transition temperatures are identical to those reported so far for ceramic samples. This matching will allow to complete the existing body of experimental information on ceramic samples (used in theoretical studies during the last 20 years) with novel, but perfectly compatible data. Complex measurements never conducted to date will become now possible, as long as the employed techniques are compatible with the crystal size. This includes Focused Ion Beam techniques, which could be used in combination with transport and magnetic measurements to investigate the anisotropy of basic physical properties, presently unknown for the full nickelate family. X-ray spectroscopies such as Angle-Resolved X-ray Photoemission or Inelastic X-ray Scattering (resonant and non-resonant) could take advantage of the tiny synchrotron X-ray beam cross sections to probe the electronic, phonon and magnon dispersion curves. Such measurements, out of the scope of this work, should provide crucial information for theoreticians and help to improve the models, fundamental energies and methods for describing the complex physics of these materials. It should also contribute to a better understanding of the complex region at the boundary between localized and itinerant behavior, and to the elaboration of design strategies, which might lead to novel transition metal oxides with improved, societally relevant functional properties.

ASSOCIATED CONTENT

**Supporting Information**

Materials, crystallographic methods, Scanning Electron Microscopy, magnetization, heat capacity, differential scanning calorimetry, phase identification for YbNiO₃ and TmiO₃, optical microscope images, REClO impurity analysis, Elemental analysis, SmClO recrystallization, Rietveld fits, RT crystal structures, tables with lattice parameters, atomic positions, reliability factors, angles, tilts, distances, Néel and Metal-Insulator transition temperatures, Curie constants, effective magnetic moments and Curie-Weiss temperatures.

**Accesion codes**

CCDC 2077728-2077729 and 2077640-2077642 contain the supplementary crystallographic data for this paper.

AUTHOR INFORMATION

**Author Contributions**

D.J.G.: Synthesized the all samples, except EuNiO3 and PrNiO3, synthesized respectively by P.L and Y.M.K.; M.K. measured the SEM and provided the EDX analysis; D.J.G, Y.M.K and M.M. conducted physical properties and PXRD measurements and analyzed the data; D.J.G, Y.M.K and T.L. measured, solved and refined the single crystal structures; M.M. wrote the manuscript with input from all authors. All authors have given approval to the final version of the manuscript.

**Funding Sources**

This work was supported by the Swiss National Science Foundation through the NCCR MARVEL (Grant No. 51NF40-182892), and the R'equip Grant n. 461 206021_139082.

ACKNOWLEDGMENTS

We thank K. Conder, J. Karpinski, A. Hampel, C. Ederer, N. Spaldin, A. Georges, O. Peil, D. van der Marel, J. Tessier, I. Ardizzone, Philip Moll and D. Mazzone for fruitful discussions.